# Investigation of the deposition of α-tantalum (110) films on a-plane sapphire substrate by molecular beam epitaxy for superconducting circuit

Haolin Jia[1], Boyi zhou[1], Tao Wang[1], Yanfu Wu[1], lina Yang[1], Zengqian Ding[1], Shuming Li[1], Xiao Cai[1], Kanglin Xiong[1,2,a)], Jiagui Feng[1,2,b)]

[1] Gusu Laboratory of Materials, Suzhou, 215123, China
[2] Vacuum Interconnected Nanotech Workstation (Nano-X), Suzhou Institute of Nano-Tech and Nano-Bionics, CAS, Suzhou, 215123, China
[a)] E-mail: klxiong2008@sinano.ac.cn
[b)] E-mail: jgfeng2017@sinano.ac.cn

## ABSTRACT

Polycrystalline α-tantalum (110) films deposited on c-plane sapphire substrate by sputtering are used in superconducting qubits nowadays. However, these films always occasionally form other structures, such as α-tantalum (111) grains and β-tantalum grains. To improve the film quality, we investigate the growth of α-tantalum (110) films on a-plane sapphire substrate under varying conditions by molecular beam epitaxy technology. The optimized α-tantalum (110) film is single crystal, with a smooth surface and atomically flat metal-substrate interface. The film with thickness of 30 nm shows a $T_c$ of 4.12K and a high residual resistance ratio of 9.53. The quarter wavelength coplanar waveguide resonators fabricated with the 150 nm optimized α-tantalum (110) film, exhibits intrinsic quality factor of over one million under single photon excitation at millikelvin temperature.

# I. INTRODUCTION

Superconducting coplanar waveguide (SCPW) resonators with low microwave loss are critical elements for quantum computation[1-4], quantum memories[5, 6] and photon detection[7]. Two-level system (TLS) is the most prominent source of dielectric loss quantized by intrinsic (unloaded) quality factor ($Q_i$) in superconducting circuits[8-10]. According to recent research, the unexplained TLS defects are mainly distributed at substrate-air, metal-air and metal-substrate interfaces[11-13]. Therefore, it is very important to improve the material growth and fabrication process to decrease the volume of these interfaces for high quality devices.

Aluminum, niobium and tantalum are mostly used in superconducting circuits[14]. Due to its high superconducting transition temperature and relatively low microwave loss, niobium (Nb) has become more commonly used materials for superconducting circuits[15]. However, the complicated stoichiometry of the native Nb oxide layer is considered to be a major source of complex microwave loss at the metal-air interface[16, 17]. In contrast, the surface insulating oxide of tantalum (Ta) is expected to be less lossy in superconducting circuits[18, 19]. Record coherence times for two-dimensional Transmon qubits have been reported by using the capacitor and microwave resonators fabricated from magnetron sputtered Ta films[20, 21]. However, these films always contain other structures, such as β -Ta grains and α-Ta (111) grains, which are harmful to high quality superconducting circuit fabrication. Additionally, comparing with the sputtering method, which is known to damage the substrate surface by ions[11] and incorporate contaminants, the molecular beam epitaxy (MBE) technology can keep the surface intact, grow high purity film, and give atomically-flat metal-air and metal-substrate interfaces. Additionally, Ta films grown by MBE usually have a more continuous structure than Ta films with a columnar structure deposited by magnetron sputtering. It is expected that Ta films grown by MBE exhibit very large grain surface area and less grain boundaries, which are related to high residual resistance ratio (RRR) and high superconducting transition temperature $T_c$[17].

In this work, to optimize the MBE growth conditions of α-Ta (110) films on a-plane (11-20) sapphire substrate, the influences of growth rate and growth temperature have been investigated by using atomic force microscopy (AFM), high-resolution X-ray diffraction (HR-XRD), and scanning transmission electron microscope (STEM). The optimized α-Ta (110) film exhibits a single-crystal structure, with a polished surface and an atomically flat metal-substrate interface. The film with thickness of 30 nm shows a $T_c$ of 4.12K and a high RRR of 9.53. The resonators of quarter-wavelength CPW produced by using the 150 nm optimized α-Ta (110) film, demonstrate an intrinsic quality factor exceeding one million under single photon excitation at millikelvin temperature.

## II. EXPERIMENTAL

The Ta films for the resonators were grown on the a-plane sapphire substrate at high temperature (500 °C ~ 800 °C) by MBE. The Ta material was evaporated through a scanning electron beam heater. The beam energy is 8KeV and the emission current was around 200mA. The deposition rate was verified using two methods, namely Atomic Absorption Spectroscopy (AAS) and Quartz Crystal Microbalance (QCM). To ensure a stable deposition rate, the signal from AAS was used as feedback to control the heating power. A high-purity Ta rod with a purity of 99.998% was used as the source material, tailored to fit the crucible shape. Before the deposition, the source material was degassed sufficiently before the deposition. Prior to the growth, two out-gassing bakes were performed for thermal cleaning of the substrate surface at 200 °C for 2 h and followed by 850 °C for 0.5 h to remove surface contamination. The growth chamber had an actual pressure of $2\times10^{-9}$ Torr during growth with the Ta electron-beam hearth operational. The surface morphology and structural properties of these Ta films were characterized using several methods including AFM, STEM, and HR-XRD.

The resonator patterns were defined on Ta film on a 2-inch sapphire wafer by optical lithography, and the Ta film was then etched with an etchant composed of 1:1:1 HF: $HNO_3$: $H_2O$. After etching, the wafer was cut into 8 mm × 8 mm square chips, and the chips were cleaned for 10 minutes in an ultrasonic bath of acetone and iso-propyl alcohol to remove particles on the samples and strip the protective photoresist. The samples are placed in an aluminum box and wired bonded, and then placed inside a dilution fridge with base temperature of 10 mK. Cryoperm shield is used to protect the devices from external magnetic fields. Filters are used to block infrared radiation. The measurement is done by a vector network analyzer (VNA) with variable output power. The photon number in the CPW can be estimated from the input power of the pump using the following equation[22]:

$$< n > \approx (Q^2 \cdot P_{in})/(\pi^2 \hbar f^2 \cdot Q_c)$$

where $n$ is the estimated photon number, $P_{in}$ is the pump power, $h$ is Planck's constant, $f$ is the resonance frequency of the CPW resonator, $Q$ is the loaded (total) quality factor of the resonator, and $Q_c$ is the value which is related to the coupling capacitance between the CPW and driveline.

## III. RESULTS AND DISCUSSION

Sapphire is a common substrate for superconducting quantum devices. The crystallographic structure and growth orientation of epitaxial Ta films are strongly affected by the substrate. The epitaxy of bcc Ta (110) can be realized on a-plane sapphire substrate due to the alignment of the 3-fold axes in the plane of the metal-substrate interface[23]. The different growth rates lead to various morphologies of the 30-nm-thick Ta (110) films grown by MBE, as shown in FIG. 1(a-d). These AFM images demonstrate the smooth surface of Ta (110) films grown with different growth rates at 550 °C for a scanning area of 1 μm × 1 μm. The measured RMS roughness values of the surface over 1 μm × 1 μm scans are about 0.15 nm. With growth rate of 0.15 A/s, the surface exhibits many pits caused by the appearance of small crystal nucleus and surface defects of substrate, which are diminished as the growth rate increases. This suggests that Ta adatoms with reduced mobility tend to stay near the point of impact, which may not be the minimum free energy lattice sites, resulting in the decrease of pits on surface. However, too large growth rates lead to the appearance of pellets on the surface of Ta film, which may be due to excessively low mobility of Ta atoms on the growth surface.

AFM images of Ta films with growth rate of 0.22 A/s at 600 °C and 800 °C are shown in FIG. 1(e) and (f), respectively. The 30-nm-thick Ta film grown at 800 °C is in the early stage of coalescence, which is formed by small islands. While, the surfaces of films grown at 500 °C and 600 °C are more continuous than that of the Ta film grown at 800 °C. The measured RMS roughness values of these surface grown at 800 °C and 600 °C over the 1 μm × 1 μm scans are 4.66 nm and 0.28 nm, respectively. These results indicate that the coalescence of Ta films slows down at higher temperature, which may be finished within several atomic layers during low temperatures growth. Higher growth temperatures promote three-dimensional growth instead of lateral growth, resulting in grooved morphologies on the surface, due to Ehrlich-Schwoebel (ES) barrier which makes unbalance between descending steps and ascending steps of Ta adatoms[24]. As mentioned above, the reduced mobility of Ta adatoms by increasing growth rate will suppress ascending steps process and promote lateral growth. Therefore, to match a higher growth temperature, a higher growth rate is required.

After mutually optimized the temperature and growth rate, FIG. 2(a) and (c) show AFM images of 150-nm-thick Ta film grown with growth rate of 0.3 A/s at 550 °C and 0.5 A/s at 800 °C, respectively. The measured RMS roughness values in FIG. 2(a) and (c) are 0.21 nm and 0.33 nm for a scanning area of 1μm × 1μm. The surface grown at 550 °C is smoother, and the atomic steps can be clearly observed from AFM images. In FIG. 2(b) and (d), HRXRD spectrum of Ta films on sapphire shows clear peaks corresponding to α-Ta (110) and sapphire (11-20). The peak width of Ta (110) grown at 800 °C is narrower than that of Ta (110) grown at 550 °C. In addition, the peak position of α-Ta (110) grown at 800 °C is further away from the peak of sapphire (11-20). These results suggest that the epitaxial strain is more relaxed at high growth temperature. It is noteworthy that the diffraction peak of Ta film grown at 550 °C has shoulder on the left. It can be inferred that the epitaxial strain in Ta film grown at low temperature is inhomogeneous, while at high temperatures strain is relaxed more rapidly by formation of misfit dislocation and coalescence of islands. Due to the merging of neighboring islands attracted to each other, it is considered that the formation of grain boundary gives rise to tensile stress which leads to the reduction of compressive stress in Ta film[25].

As mentioned above, the microwave loss is mainly from the amorphous oxide at surfaces and interfaces of superconducting quantum devices[11-13]. The exposed grain boundaries will deepen the oxidization of the superconductors. In other words, the more grain boundaries, the larger microwave loss volume[17]. A high quality and structurally continuous film portends good device performance. Therefore, interface atomic structure and performance characterization of Ta film grown with optimized growth condition (0.3 A/s at 550 °C) was conducted. As shown in FIG. 3 (a)-(c), STEM of Ta film cross section indeed reveals a continuous structure, with no obvious grain boundary described as Josephson weak links[26, 27]. The growth direction is oriented along the [110] axis (FIG. 3(a)). It can be seen that the amorphous tantalum oxide layer is about 2 nm thick on the top of Ta (FIG. 3(b)), with clear interface between Ta and the oxide layer. The interface between Ta and sapphire is atomically flat and the arrangement pattern of Ta atoms in the film continues the arrangement pattern of Al atoms in the substrate (The inset of FIG.3(c)), showing an epitaxial growth. In the interfacial region, there is no misfit dislocation found from the STEM image in FIG. 3(c). Finally, combining AFM studies of Ta surface and cross section from STEM image, it indicates that the Ta films grows layer by layer at 550 °C with proper growth rates, avoiding defects caused by the coalescence of 3D islands. In FIG. 3(d), the Ta (101)-diffraction peak series show four peaks and indicate no twist domains in the Ta film. This is consistent with the analysis of the interface atomic structure of Ta (110) film epitaxy on a-plane sapphire substrate.

The observed superconducting transition temperature $T_c$ of the Ta film (30-nm-thick) is 4.12 K and RRR is 9.53 as shown in FIG. 4(a), further conforming the high quality of the Ta film. Then, 150-nm-thick Ta film grown with growth rate of 0.3 A/s at 550 °C was patterned into SCPW microwave resonator in the form of a quarter-wavelength segment to extract $Q_i$ from the $S_{21}$ transmission measurement. The structure of the resonator is shown in Fig. S1 (a) in the supplementary material [URL will be inserted by AIP Publishing]. The $Q_i$ was extracted by fitting the S21 vs. frequency curve. One fitting example is shown in Fig. S1 (b) and (c) in the supplementary material [URL will be inserted by AIP Publishing].The device was designed with center conductor and insulating gap widths of $w$ = 10 μm and $g$ = 6 μm, respectively. Resonance frequencies ***f*** ranged from 5 to 7 GHz. FIG. 4(b) shows the dependence of $Q_i$ on the microwave drive power, all resonators with different frequency show $Q_i$ value higher than one million under single photon region. The $Q_i$ shows an increasing trend with the increase of power. As the input power increase further, obvious nonlinear effects occur, see Fig. S2 in the supplementary material [URL will be inserted by AIP Publishing]. It is clear that the high $Q_i$ resonator made from Ta film grown by MBE has been realized, due to the atomically flat interface between Ta and sapphire, and the smooth surface of high-quality Ta film with single crystal structure.

It is worth noting that the etching rate of Ta film grown by MBE is significantly slower than that of Ta film deposited by magnetron sputtering due to the better compactness of the film, under-etching phenomenon occurred during the preparation process of the CPW microwave resonator. The $Q_i$ value of the resonator made from MBE-grown Ta film is not higher than some resonators made from deposited Ta films by magnetron sputtering in other works, which could be caused by the processing technique of the (a-plane) sapphire and the etching process of CPW. In other words, both the a-plane sapphire substrate preparation and the repeatability of the optimal etching process for the CPW need to be improved to adapt to MBE-grown Ta films and further research is needed to identify other loss channels.

## IV. CONCLUSIONS

In conclusion, we optimized the MBE growth conditions of α-Ta (110) films on a-plane sapphire substrates. The effects of growth rate and temperature on the MBE-grown film properties were investigated using various characterization tools, such as AFM, HR-XRD, and STEM. The optimized α-Ta (110) film exhibited a single-crystal structure with a polished surface and an atomic metal-substrate interface. The film with thickness of 30 nm shows a $T_c$ of 4.12K and a high RRR of 9.53. Resonators based on quarter-wavelength CPW were fabricated by utilizing the 150 nm optimized α-Ta (110) film, which demonstrated $Q_i$ exceeding one million when subjected to single photon excitation at millikelvin temperature.

## ACKNOWLEDGMENTS

K. L. X acknowledges support from the Youth Innovation Promotion Association of Chinese Academy of Sciences (2019319). J. G. F. acknowledges support from the Start-up foundation of Suzhou Institute of Nano-Tech and Nano-Bionics, CAS, Suzhou (Y9AAD110).

## CONFLICT OF INTEREST

The authors have no conflicts to disclose.

## DATA AVAILABILITY

The data that support the findings of this study are available from the corresponding author upon reasonable request.

## REFERENCES

1. C. J. K. Richardson, A. Alexander, C. G. Weddle, B. Arey and M. Olszta, Journal of Applied Physics **127** (23), 235302 (2020).
2. H.-L. Huang, D. Wu, D. Fan and X. Zhu, Science China Information Sciences **63** (8), 180501 (2020).
3. S. Kwon, A. Fadavi Roudsari, O. W. B. Benningshof, Y.-C. Tang, H. R. Mohebbi, I. A. J. Taminiau, D. Langenberg, S. Lee, G. Nichols, D. G. Cory and G.-X. Miao, Journal of Applied Physics **124** (3), 033903 (2018).
4. P. Jurcevic, A. Javadi-Abhari, L. S. Bishop, I. Lauer, D. F. Bogorin, M. Brink, L. Capelluto, O. Günlük, T. Itoko and N. Kanazawa, Quantum Science and Technology **6** (2), 025020 (2021).
5. K. Sardashti, M. C. Dartiailh, J. Yuan, S. Hart, P. Gumann and J. Shabani, IEEE Transactions on Quantum Engineering **1**, 1-7 (2020).
6. M. Hofheinz, H. Wang, M. Ansmann, R. C. Bialczak, E. Lucero, M. Neeley, A. D. O'Connell, D. Sank, J. Wenner, J. M. Martinis and A. N. Cleland, Nature **459** (7246), 546-549 (2009).
7. P. J. De Visser, S. A. H. De Rooij, V. Murugesan, D. J. Thoen and J. J. A. Baselmans, Physical Review Applied **16** (3), 034051 (2021).

8. D. Niepce, J. J. Burnett, M. G. Latorre and J. Bylander, Superconductor Science and Technology **33** (2), 025013 (2020).
9. G. Calusine, A. Melville, W. Woods, R. Das, C. Stull, V. Bolkhovsky, D. Braje, D. Hover, D. K. Kim, X. Miloshi, D. Rosenberg, A. Sevi, J. L. Yoder, E. Dauler and W. D. Oliver, Applied Physics Letters **112** (6), 062601 (2018).
10. J. M. Sage, V. Bolkhovsky, W. D. Oliver, B. Turek and P. B. Welander, Journal of Applied Physics **109** (6), 063915 (2011).
11. A. Megrant, C. Neill, R. Barends, B. Chiaro, Y. Chen, L. Feigl, J. Kelly, E. Lucero, M. Mariantoni, P. J. J. O'Malley, D. Sank, A. Vainsencher, J. Wenner, T. C. White, Y. Yin, J. Zhao, C. J. Palmstrøm, J. M. Martinis and A. N. Cleland, Applied Physics Letters **100** (11), 113510 (2012).
12. J. Wenner, R. Barends, R. C. Bialczak, Y. Chen, J. Kelly, E. Lucero, M. Mariantoni, A. Megrant, P. J. J. O'Malley, D. Sank, A. Vainsencher, H. Wang, T. C. White, Y. Yin, J. Zhao, A. N. Cleland and J. M. Martinis, Applied Physics Letters **99** (11), 113513 (2011).
13. A. Bilmes, A. Megrant, P. Klimov, G. Weiss, J. M. Martinis, A. V. Ustinov and J. Lisenfeld, Sci Rep **10** (1), 3090 (2020).
14. Xiong K L, Feng J G, Zhen Y R, Cui J Y, Weng W K, Zhang S Y, Li S F and Y. H., Chin Sci Bull **67**, 143–162 (2022).
15. M. S. Blok, V. V. Ramasesh, T. Schuster, K. O'Brien, J. M. Kreikebaum, D. Dahlen, A. Morvan, B. Yoshida, N. Y. Yao and I. Siddiqi, Physical Review X **11** (2), 021010 (2021).
16. D. Bach, H. Stormer, R. Schneider, D. Gerthsen and J. Verbeeck, Microsc Microanal **12** (5), 416-423 (2006).
17. A. Premkumar, C. Weiland, S. Hwang, B. Jäck, A. P. M. Place, I. Waluyo, A. Hunt, V. Bisogni, J. Pelliciari, A. Barbour, M. S. Miller, P. Russo, F. Camino, K. Kisslinger, X. Tong, M. S. Hybertsen, A. A. Houck and I. Jarrige, Communications Materials **2** (1) (2021).
18. R. A. McLellan, A. Dutta, C. Zhou, Y. Jia, C. Weiland, X. Gui, A. P. M. Place, K. D. Crowley, X. H. Le, T. Madhavan, Y. Gang, L. Baker, A. R. Head, I. Waluyo, R. Li, K. Kisslinger, A. Hunt, I. Jarrige, S. A. Lyon, A. M. Barbour, R. J. Cava, A. A. Houck, S. L. Hulbert, M. Liu, A. L. Walter and N. P. de Leon, Adv Sci (Weinh), e2300921 (2023).
19. Kevin D. Crowley, Russell A. McLellan, Aveek Dutta, Nana Shumiya, Alexander P. M. Place, Xuan Hoang Le, T. M. Youqi Gang, Nishaad Khedkar, Yiming Cady Feng, Esha A. Umbarkar, Xin Gui, Lila V. H. Rodgers, Yichen Jia, Mayer M. Feldman, Stephen A. Lyon, Mingzhao Liu, Robert J. Cava, Andrew A. Houck and N. P. d. Leon, arXiv preprint arXiv **2301**, 07848 (2023).
20. C. Wang, X. Li, H. Xu, Z. Li, J. Wang, Z. Yang, Z. Mi, X. Liang, T. Su, C. Yang, G. Wang, W. Wang, Y. Li, M. Chen, C. Li, K. Linghu, J. Han, Y. Zhang, Y. Feng, Y. Song, T. Ma, J. Zhang, R. Wang, P. Zhao, W. Liu, G. Xue, Y. Jin and H. Yu, npj Quantum Information **8** (1), 3 (2022).
21. A. P. M. Place, L. V. H. Rodgers, P. Mundada, B. M. Smitham, M. Fitzpatrick, Z. Leng, A. Premkumar, J. Bryon, A. Vrajitoarea, S. Sussman, G. Cheng, T. Madhavan, H. K. Babla, X. H. Le, Y. Gang, B. Jack, A. Gyenis, N. Yao, R. J. Cava, N. P. de Leon and A. A. Houck, Nat Commun **12** (1), 1779 (2021).
22. S. Geaney, D. Cox, T. Honigl-Decrinis, R. Shaikhaidarov, S. E. Kubatkin, T. Lindstrom, A. V. Danilov and S. E. de Graaf, Sci Rep **9** (1), 12539 (2019).
23. Ward R, Grier E J and P.-L. A. K, Journal of Materials Science: Materials in Electronics **14** (9), 533-539 (2003).
24. F. F. Leal, S. C. Ferreira and S. O. Ferreira, J Phys Condens Matter **23** (29), 292201 (2011).
25. E. Chason, J. W. Shin, S. J. Hearne and L. B. Freund, Journal of Applied Physics **111** (8), 083520

(2012).
26. C. Attanasio, L. Maritato and R. Vaglio, Physical Review B **43** (7), 6128 (1991).
27. R. L. Peterson and J. W. Ekin, Physical Review B **42** (13), 8014 (1990).

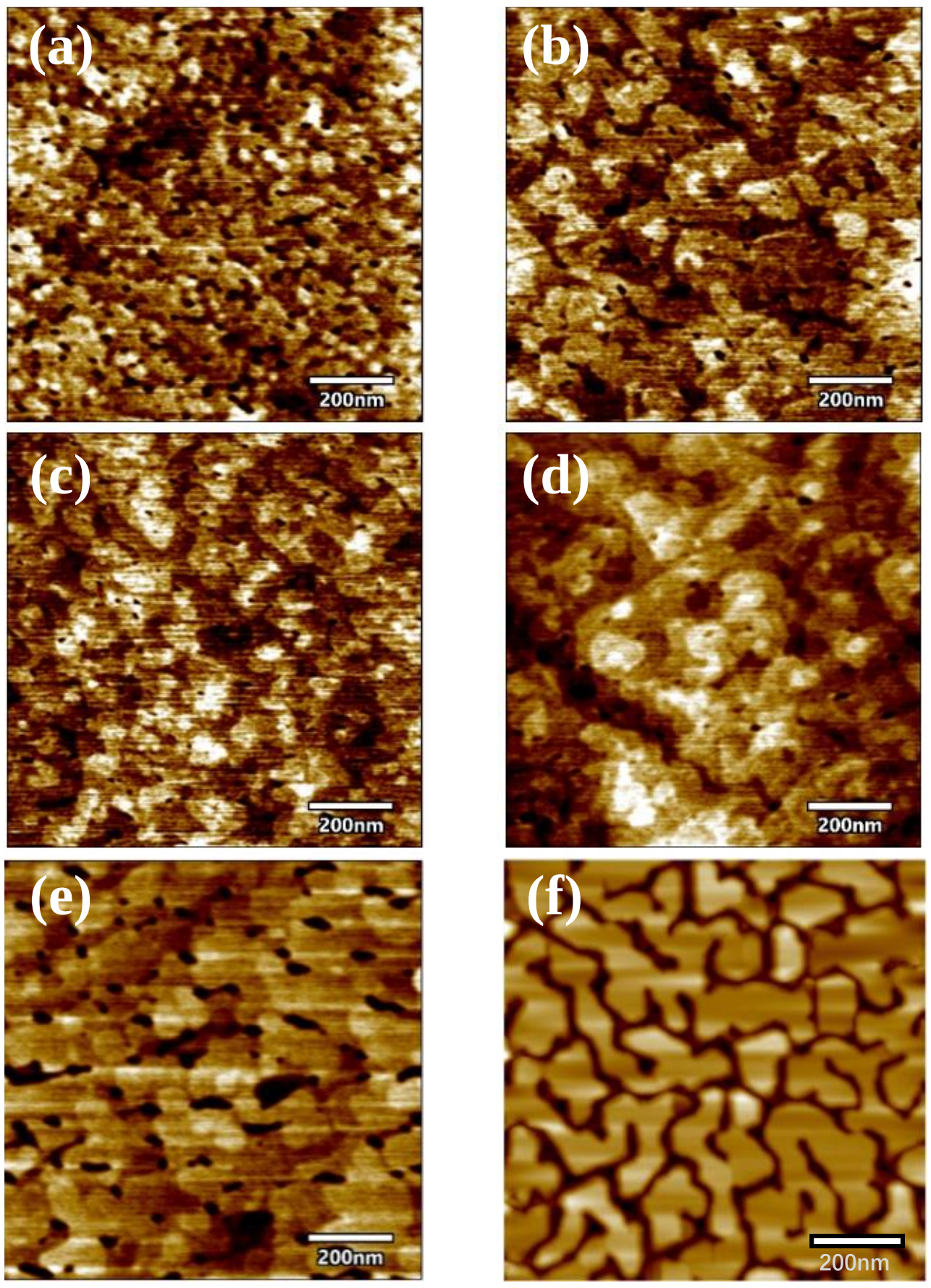

FIG. 1. (a)-(d) are AFM images of 30-nm-thick Ta (110) films with the corresponding growth rates of 0.15A/s, 0.22A/s, 0.26A/s and 0.30A/s at 550°C. (e) and (f) are 30-nm-thick Ta (110) films with growth rate of 0.22 A/s at 600°C and 800°C respectively.

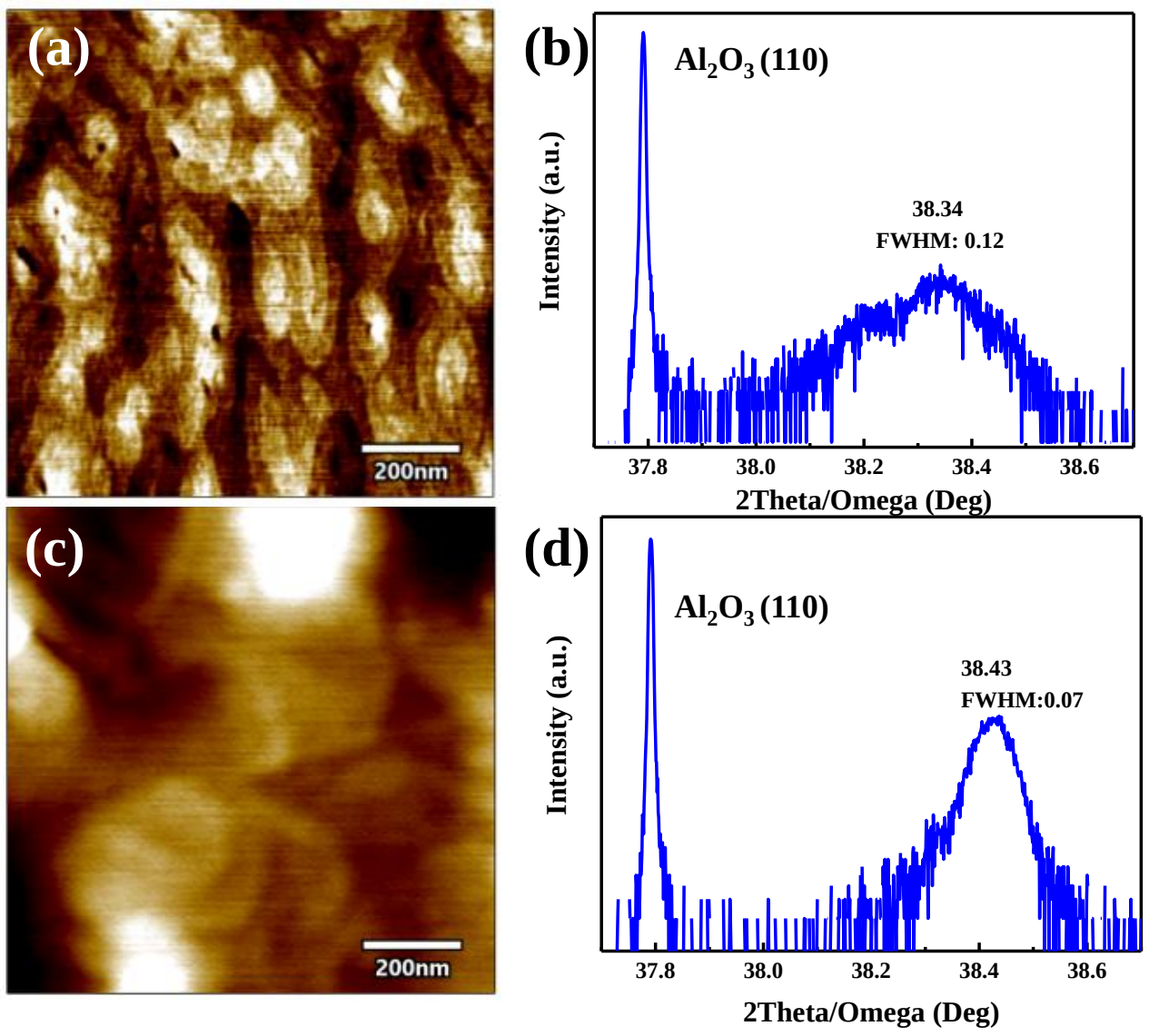

FIG. 2. (a) AFM images and (b) x-ray diffraction spectrum of Ta film grown with growth rate of 0.3 A/s at 550 °C, (c) AFM images and (d) x-ray diffraction scan of Ta film grown with growth rate of 0.5 A/s at 800 °C.

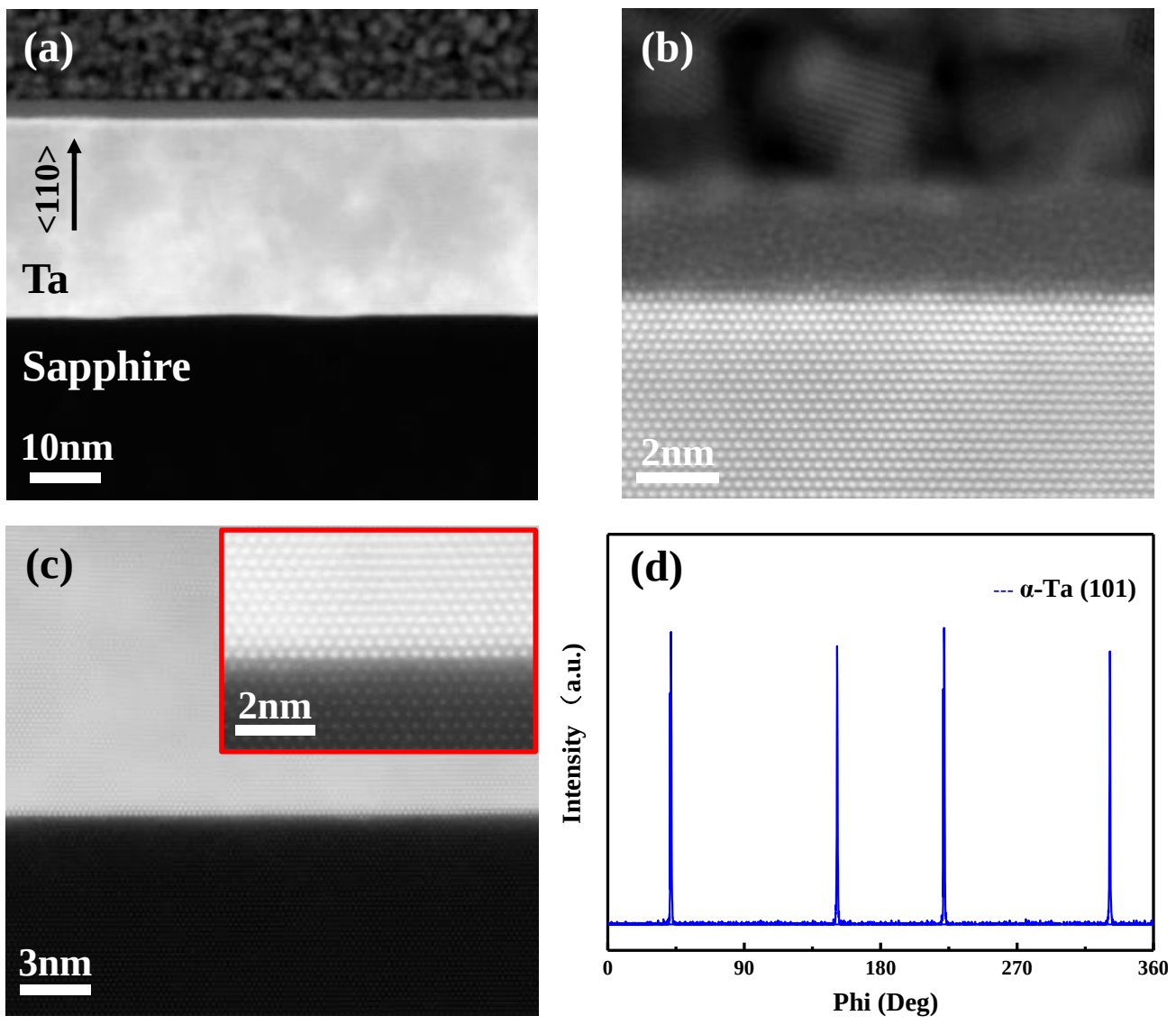

FIG. 3. (a) STEM image of the Ta (110) film, Atomic-resolution STEM image of (b) the interface between tantalum and surface oxide layer. (c) The interface between tantalum and sapphire showing epitaxial growth, and the inset shows that the arrangement pattern of Ta atoms in the film continues the arrangement pattern of Al atoms in the substrate. (d) Phi scans of the Ta (101)-diffraction peak series.

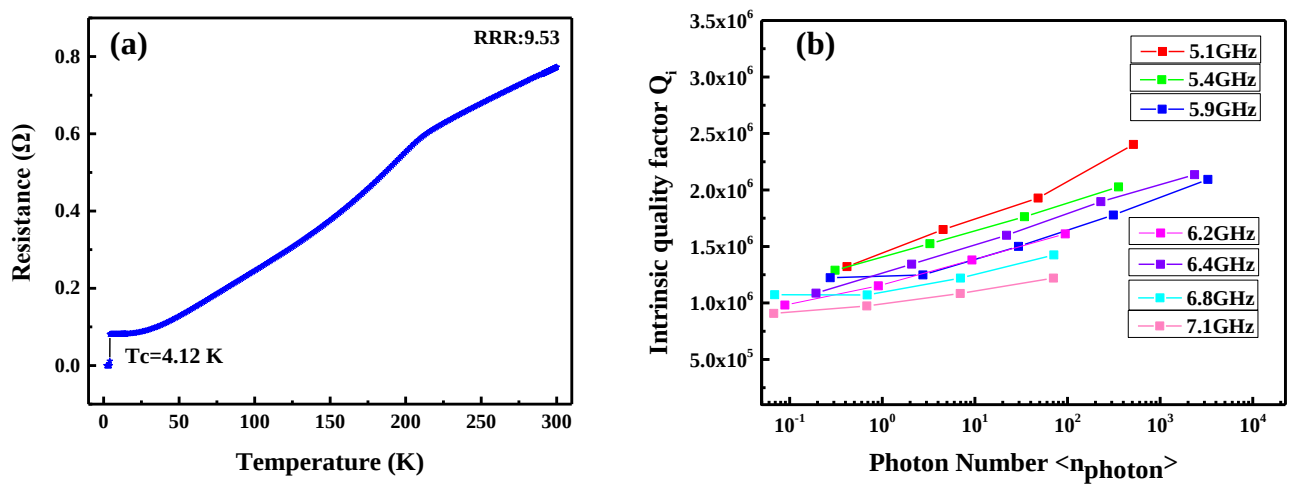

FIG. 4. (a) Four-probe resistance measurement of 30-nm-thick Ta film grown by MBE showing $T_c$ = 4.12 K, (b) The dependence of $Q_i$ on the microwave drive power for the device made from the Ta film grown with growth rate of 0.3 A/s at 550 °C.